\documentclass[showpacs,aps,prl,twocolumn,floatfix,superscriptaddress]{revtex4}
\usepackage{graphicx,graphics}
\begin {document}
\title{\bf{Effects of Bulk Viscosity in Non-linear Bubble Dynamics}}
\author{A. Moshaii}
\email{moshaii@mehr.sharif.edu} \affiliation{Department of
Physics, Sharif University of Technology, P.O. Box:11365-9161,
Tehran, I.R. Iran.} \affiliation{Institute for Studies in
Theoretical Physics and Mathematics, P.O. Box:19395-5531, Tehran,
I.R. Iran.}
\author{R. Sadighi-Bonabi}
\affiliation{Department of Physics, Sharif University of
Technology, P.O. Box:11365-9161, Tehran, I.R. Iran.}
\affiliation{Bonab Research Center, P.O. Box:56515-196, Bonab,
Azarbayejan Province, I.R. Iran.}
\author{M. Taeibi-Rahni}
\affiliation{Department of Aerospace Engineering, Sharif
University of Technology, P.O. Box:11365-9161, Tehran, I.R. Iran}
\pacs{47.55.Bx, 43.25.Yw, 43.25.+y, 78.60.Mq}

\begin{abstract}

The non-linear bubble dynamics equations in a compressible liquid
have been modified considering the effects of compressibility of
both the liquid and the gas at the bubble interface. A new bubble
boundary equation has been derived, which includes a new term
resulted from the liquid bulk viscosity effects. The influence of
this term has been numerically investigated considering the
effects of water vapor and chemical reactions on the bubble
evolution. The results clearly indicate that the new term has an
important damping role at the collapse, so that its consideration
decreases the amplitude of the bubble rebounds after the
collapse. This damping feature is more remarkable for higher
deriving pressures.
\end{abstract}
\maketitle

When a small isolated gas bubble, immersed in a liquid,
experiences a high amplitude spherical sound field, it grows and
contracts non-linearly. Description of the dynamics of such
non-linear motion is an old challenging problem. The radial
dynamics of the bubble in an incompressible liquid is described by
the well-known incompressible Rayleigh-Plesset equation
\cite{Rayleigh:1917,Noltingk:1951}. The extension of this equation
to the bubble motion in a compressible liquid has been studied by
many previous authors \cite{Herring:1941, Keller:1980}. The most
complete existing description was presented by Prosperetti and
Lezzi \cite{Prosperetti:1987}. They used a singular-perturbation
method of the bubble-wall Mach number and derived a one-parameter
family of equations describing the bubble motion in the first
order approximation of compressibility. This family of equations
are written as:

\begin{eqnarray}
\label{eq1} \left({1-(\eta + 1)\frac{\dot{R}}{C}}
\right)\!R\ddot{R}+\frac{3}{2}\left({1-\frac{1}{3}(3\eta +
1)\frac{\dot {R}}{C}}\right)\!\dot{R}^2 = \nonumber \\
{\frac{R}{\rho C}\frac{d}{dt}\left( {P_l-P_a}
\right)\!+\!\left({1+(1-\eta)\frac{\dot {R}}{C}} \
\right)\!\!\!\left( \frac{P_l-P_a-P_0}{\rho}
\right)},\nonumber \\
\end{eqnarray}

\noindent where, $R$, $C$, $P_0$, $P_a$, and $\rho$ are bubble
radius, liquid sound speed, ambient pressure, driving pressure,
and density of the liquid, respectively. Also, $\eta$ is an
arbitrary parameter. Equation (\ref{eq1}) must be supplemented by
a boundary condition equation at the bubble interface to relate
the liquid pressure, $P_{l}$, to the gas pressure inside the
bubble. Like all previous authors, Prosperetti and Lezzi
\cite{Prosperetti:1987} used the following incompressible
equation for this purpose:

\begin{equation}
\label{eq2}
P_l = P_g - 4\mu \frac{\dot {R}}{R}-\frac{2\sigma}{R},
\end{equation}

\noindent where, $P_{g}$, $\mu$, and $\sigma$ are gas pressure at
the bubble interface, liquid viscosity coefficient, and surface
tension, respectively. Most of the previously obtained equations
belong to this single parameter family of equations,
corresponding to different values of $\eta$. Moreover, $\eta = 0$
yields results in closest agreement with the numerical simulation
of full partial differential equations \cite{Prosperetti:1987}.

In all previous works \cite{Rayleigh:1917, Noltingk:1951,
Herring:1941, Keller:1980, Prosperetti:1987}, an important
approximation has been used in the derivation of the bubble
dynamics equations. That is the incompressibility assumption of
the liquid motion at the bubble interface, which has been used in
the derivation of Eq'n. (\ref{eq2}). Note that, all of the
effects of the liquid compressibility in all previous papers have
been resulted from the liquid motion around the bubble, but not
from the bubble boundary condition equation. In fact, all
previous authors, on one hand took into account the
compressibility of the liquid motion around the bubble, but on
the other hand neglected its consideration at the bubble
interface.

In this paper, we have modified the bubble dynamics equations
considering the effects of the liquid compressibility at the
bubble interface. We have derived a new bubble boundary equation
instead of Eq'n. (\ref{eq2}). The new equation has new terms
resulted from the effects of bulk viscosity of the liquid and the
gas.

To derive the compressible bubble boundary equation, the
continuity equation and the radial component of the stress tensor
under the spherical symmetric condition can be written as:
\begin{equation}
\label{eq3} \frac{1}{\rho }\left[ {\frac{\partial \rho }{\partial
t}+u\frac{\partial \rho }{\partial r}} \right] = - \frac{\partial
u}{\partial r}-\frac{2u}{r}=-\Delta,
\end{equation}

\begin{equation}
\label{eq4} T _{rr}=-p+(\mu_b-\frac{2\mu}{3})
\Delta+2\mu\left(\frac{\partial u }{\partial r}\right).
\end{equation}

\noindent where, $\rho$, $u$, $p$, and $\Delta $ are density,
velocity, pressure, and divergence of the velocity, respectively.
Also, $\mu_b$ is the bulk viscosity coefficient and is defined by
$\mu_b=\lambda+2\mu/3$, where $\lambda$ is second coefficient of
viscosity \cite{White:1991}. Inserting $\partial u/\partial r$
from Eq'n. (\ref{eq3}), into Eq'n. (\ref{eq4}) yields:
\begin{eqnarray}
\label{eq5}T_{rr}=-p+(\mu_{b}+\frac{4\mu}{3})\triangle-4\frac{\mu u}{r}.
\end{eqnarray}

\noindent The velocity divergence, $\triangle$, can be written as:
\begin{equation}
\label{eq6}
\triangle=-\frac{1}{\rho}\frac{d\rho}{dt}=-\frac{1}{\rho
c^{2}}\frac{dp}{dt},
\end{equation}

\noindent where, the sound speed, $c$, is defined as
$c^{2}=dp/d\rho$. The boundary continuity requirement at the
bubble interface is:
\begin{eqnarray}
\label{eq7} T_{rr}(liquid)\mid_{R}=
T_{rr}\left(gas\right)\mid_{R}+2\frac{\sigma}{R}.
\end{eqnarray}

\noindent Applying Eq'n. (\ref{eq5}) for the gas and the liquid
parts of Eq'n. (\ref{eq7}) leads to:
\begin{eqnarray}
\label{eq7.5} P_{l}+4\frac{\mu\dot{R}}{R} &-&
\left(\mu_{b}+\frac{4\mu}{3}\right)\triangle_{l}= P_{g}+
4\frac{\mu_{g}\dot{R}}{R}
 \nonumber \\ &-&
\left(\mu_{bg}+\frac{4\mu_{g}}{3}\right)\triangle_{g}-2\frac{\sigma}{R},
\end{eqnarray}

\noindent where, $\mu_g$ and $\mu_{bg}$ are the viscosity and the
bulk viscosity coefficients of the gas at the bubble interface,
respectively. Also, $\Delta_{l}$ and $\Delta_{g}$ are the
divergence of velocity of the liquid and the gas, respectively.
Substituting the divergence of velocity for the liquid and the
gas from Eq'n (\ref{eq6}) into Eq'n. (\ref{eq7.5}) yields:

\begin{eqnarray}
\label{eq8} P_{l}+4\frac{\mu\dot{R}}{R} &+&
\left(\frac{\mu_{b}}{\rho C^{2}}+\frac{4\mu}{3\rho
C^{2}}\right)\frac{dP_{l}}{dt}=P_{g}+4\frac{\mu_{g}\dot{R}}{R}
 \nonumber \\ &+& \left(\frac{\mu_{bg}}{\rho_{g}}+\frac{4\mu_{g}}{3\rho_{g}}\right)\frac{d
\rho_{g}}{dt}-2\frac{\sigma}{R},
\end{eqnarray}
\noindent where, $\rho_g$ is the gas density at the bubble
interface. Equation (\ref{eq8}) represents the bubble boundary
equation containing all effects of the compressibility and the
viscosity of both the liquid and the gas. Comparison of Eq'ns.
(\ref{eq2}) and (\ref{eq8}) indicates the existence of three new
terms in Eq'n. (\ref{eq8}) due to the liquid and the gas
compressibility and viscosity effects. Here, we concentrate on
the effects of the new term arising from the liquid
compressibility. Therefore, we neglect the gas viscosity because
of its smallness relative to the liquid viscosity as in previous
works
\cite{Rayleigh:1917,Noltingk:1951,Herring:1941,Keller:1980,Prosperetti:1987}.
Under this circumstance, Eq'n. (\ref{eq8}) becomes:

\begin{equation}
\label{eq9} P_{l}+\left(\frac{\mu_{b}}{\rho
C^{2}}+\frac{4\mu}{3\rho
C^{2}}\right)\frac{dP_{l}}{dt}=P_{g}-4\frac{\mu\dot{R}}{R}-2\frac{\sigma}{R}.
\end{equation}
\noindent It should be mentioned that, although the effects of
compressibility consideration in Eq'n. (\ref{eq1}) are in the
first order approximation, but these effects have been introduced
completely in Eq'n. (\ref{eq9}).

To close the mathematical analysis, the gas pressure evolution at
the bubble interface, $P_{g}$, must be specified. In the most
complete approach, it can be determined from the simultaneous
solution of the conservation equations for the bubble interior
and the bubble radius equations \cite{C.C.WU, Moss:1994,
Kondict:1995, Voung:1996, Yuan:1998, Xu:2003}. Also, heat
conduction and mass exchange between the bubble and the
surrounding liquid affect the bubble evolution. In addition,
chemical reactions occurring in the high temperature conditions
at the end of the collapse, change the bubble content
\cite{Kamath:1993, Yasui:1997}. All these complexities have been
considered in a complete gas dynamics model by Storey and Szeri
\cite{Storey:2000}.

On the other hand, strong spatial inhomogeneities inside the
bubble are not remarkably revealed, unless at the end of an
intense collapse \cite{Voung:1996, Yuan:1998}. Therefore, the
uniformity assumption for the bubble interior seems to be useful
and provides many features of the bubble motion
\cite{Barber:1997, Brenner:2002}. Using this assumption,
recently, Lohse and his coworkers presented an ODE model
\cite{Toegel:2002, Toegel:2003, Lu:2003}, in which all effects of
heat transfer at the bubble interface, phase change of water
vapor, chemical reactions, and diffusion of reaction products
have been considered. This model accurately describes various
experimental phase diagrams \cite{Toegel:2003} and provides a
good agreement with the complete direct numerical simulation of
Storey and Szeri \cite{Toegel:2002, Storey:2000}.

Here, for describing the bubble interior evolution, we have used
the Lohse's group model (the same as what has been presented in
Ref. \cite{Lu:2003}). We do not repeat this model here and for
more details we refer to Refs. \cite{Toegel:2003, Lu:2003}. The
calculations were carried out under the framework of Eq'n
(\ref{eq1}), ($\eta=0$), for both the new compressible (Eq'n.
\ref{eq9}) and the old incompressible (Eq'n. \ref{eq2}) boundary
conditions. We describe an argon bubble in water at room
temperature, $T_{0}=293.0~K$, and atmosphere pressure, $P_{0}=1.0
~atm$, under the conditions of Single Bubble Sonoluminescence
\cite{Brenner:2002, Barber:1997}. The driving pressure was
$P_{a}(t)=P_{a}\sin \left(\omega t\right)$, where
$\omega=2\pi\times26.5~kHz$. The constants and the parameters
were set accordingly \cite{CRC:1991}; $\rho=998.0 ~kg/m^{3}$,
$C=1483.0 ~m/s$, $\mu=1.01\times10^{-3} ~kg/ms$, $\sigma=0.0707
~kgs^{-2}$. The bulk viscosity of water at room temperature was
set to be $\mu_{b}=4.1\times10^{-3} ~kg/ms$ \cite{Karim:1952}.
The constants and parameters of the gas evolution model were set
the same as what has been presented in Ref. \cite{Lu:2003}.

\begin{figure}[t]
\vskip 2mm
\includegraphics[width=7.5cm,height=3.7cm]{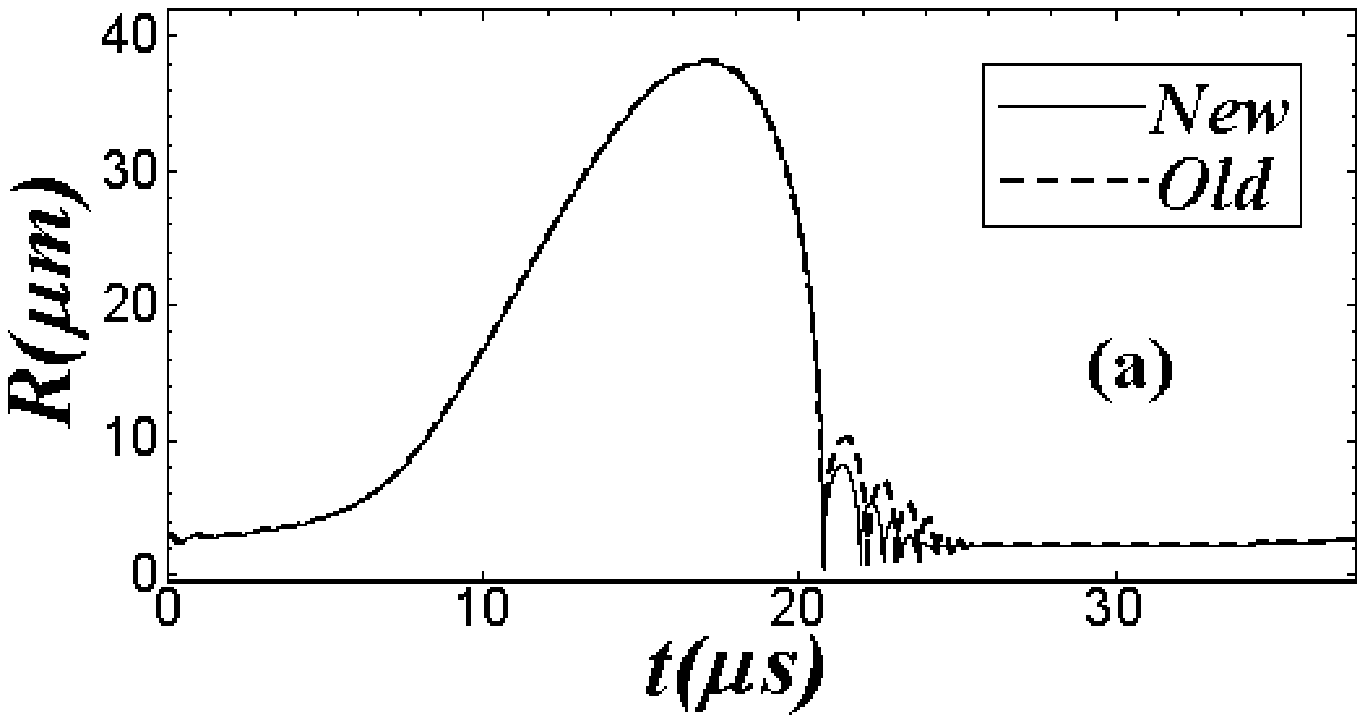}
\vskip 8mm
\includegraphics[width=7.5cm,height=3.7cm]{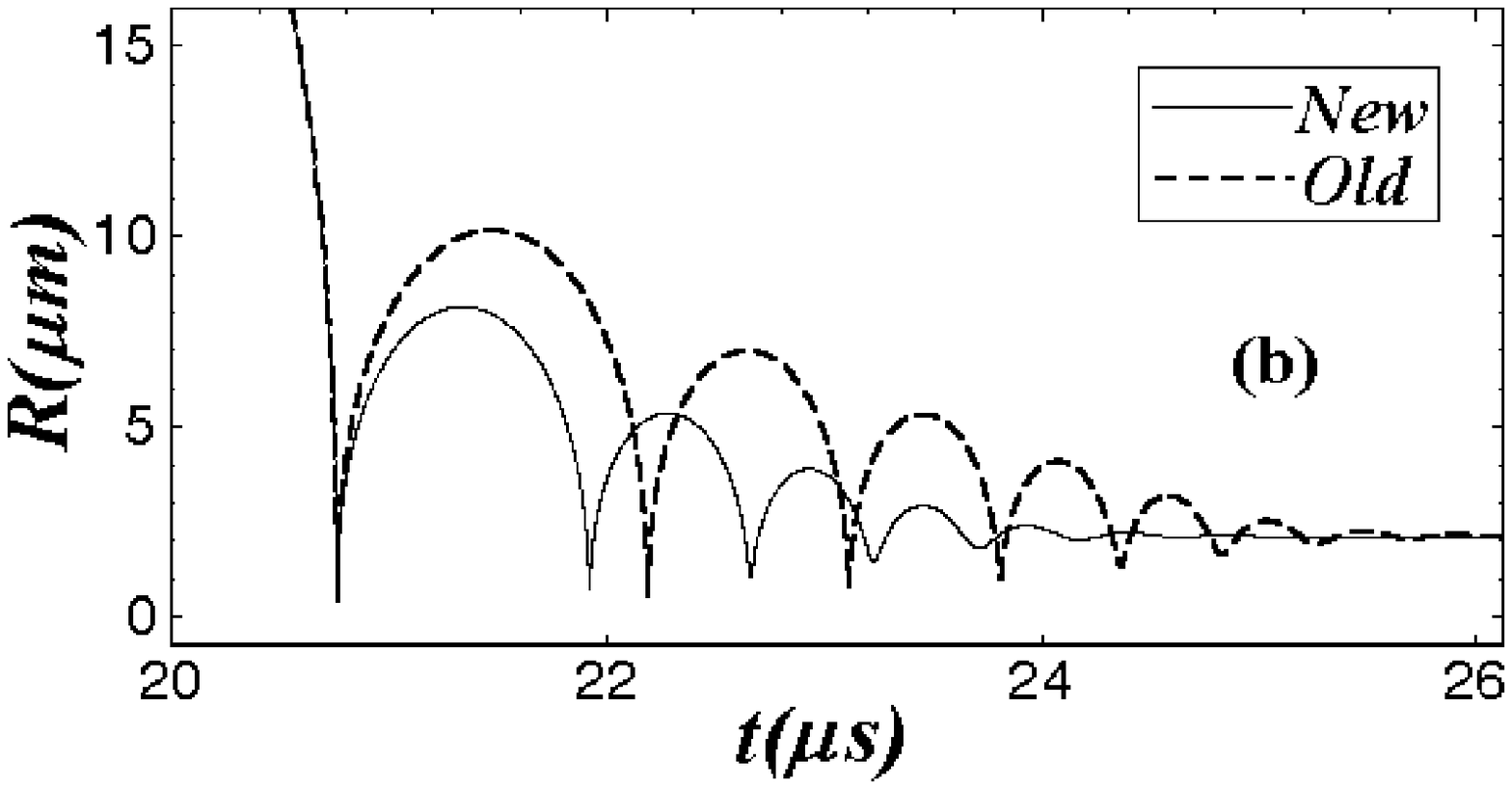}
\vskip 8mm
\includegraphics[width=7.5cm,height=3.7cm]{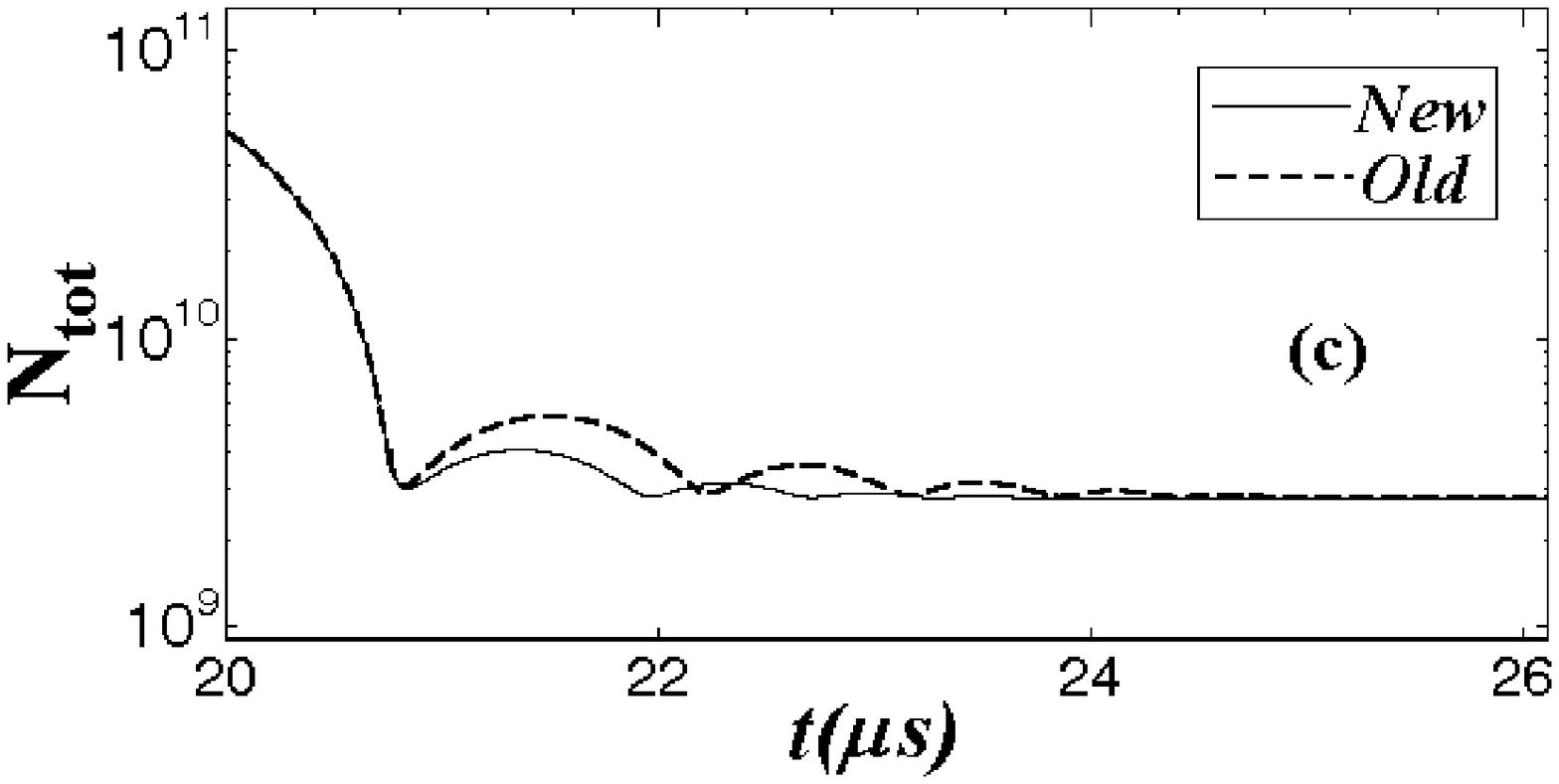}
\caption{(a) Time variations of the bubble radius in one period
according to the new compressible (solid) and the old
incompressible cases. (b) Details of the bubble rebounds after the
collapse for the two cases. (c) Variation of total number of the
particle species for the two cases during the bubble rebounds.
The equilibrium radius is $R_{0}=3.0 ~\mu m$ and the deriving
pressure is $P_{a}=1.35 ~atm$.} \label{fig1:dls}
\end{figure}
Figures (1) and (2) illustrate the variations of the bubble
characteristics (radius, total number of particle species, and
temperature), for the two boundary condition cases. It is
observed that the addition of the new viscous term in Eq'n.
(\ref{eq9}) considerably changes the bubble evolution after the
collapse. The bubble motion is remarkably compressible during the
collapse. Therefore, the new viscous term, which has been arisen
from the liquid compressibility, is important in this time
interval. This term exhibits a damping role and its consideration
reduces the amplitude of the bubble rebounds. Also, the period of
the rebounds decreases with the addition of the new term. Details
of our calculations show that the minimum radius for the new case
is about 10\% greater than that of the old one. The difference
between the two cases also appears on the variations of the total
number of particle species (Ar and H$_2$O plus reactions
products) after the collapse (Fig. 1c). Note that, the difference
gradually disappears as the bubble rebounds weaken.

\begin{figure}[t]
\vskip 2mm
\includegraphics[width=7.5cm,height=3.7cm]{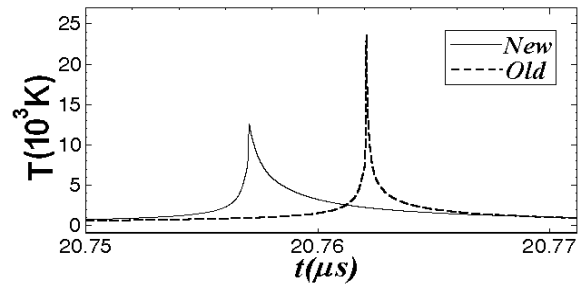}
\caption{Time variations of the gas temperature when the bubble
reaches its minimum radius according to compressible (solid) and
incompressible (dashed) boundary conditions. The parameters and
constants are the same as Fig. (1).} \label{fig3:dls}
\end{figure}
In Fig. (2), details of the gas temperature evolution around the
minimum radius have been demonstrated. Damping feature of the new
term is clearly observed by a considerable decrease of the peak
temperature (about 50\%) and an increase of the temperature pulse
width, at the collapse time. Also, the time of the peak
temperature about 5 ns changes with the addition of the new term.

\begin{figure}[t]
\vskip 2mm
\includegraphics[width=7.4cm,height=3.7cm]{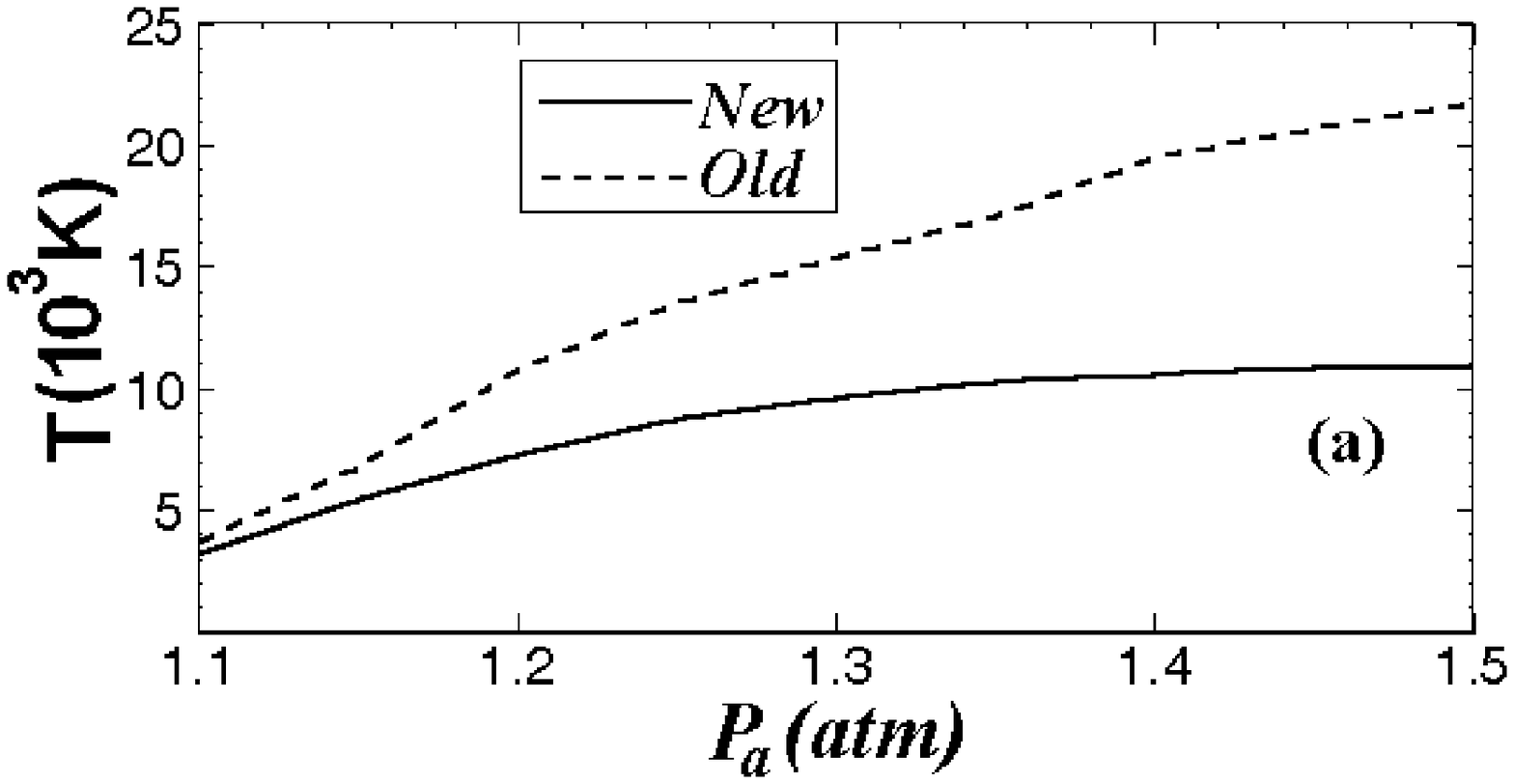}
\vskip 8mm
\includegraphics[width=7.3cm,height=3.7cm]{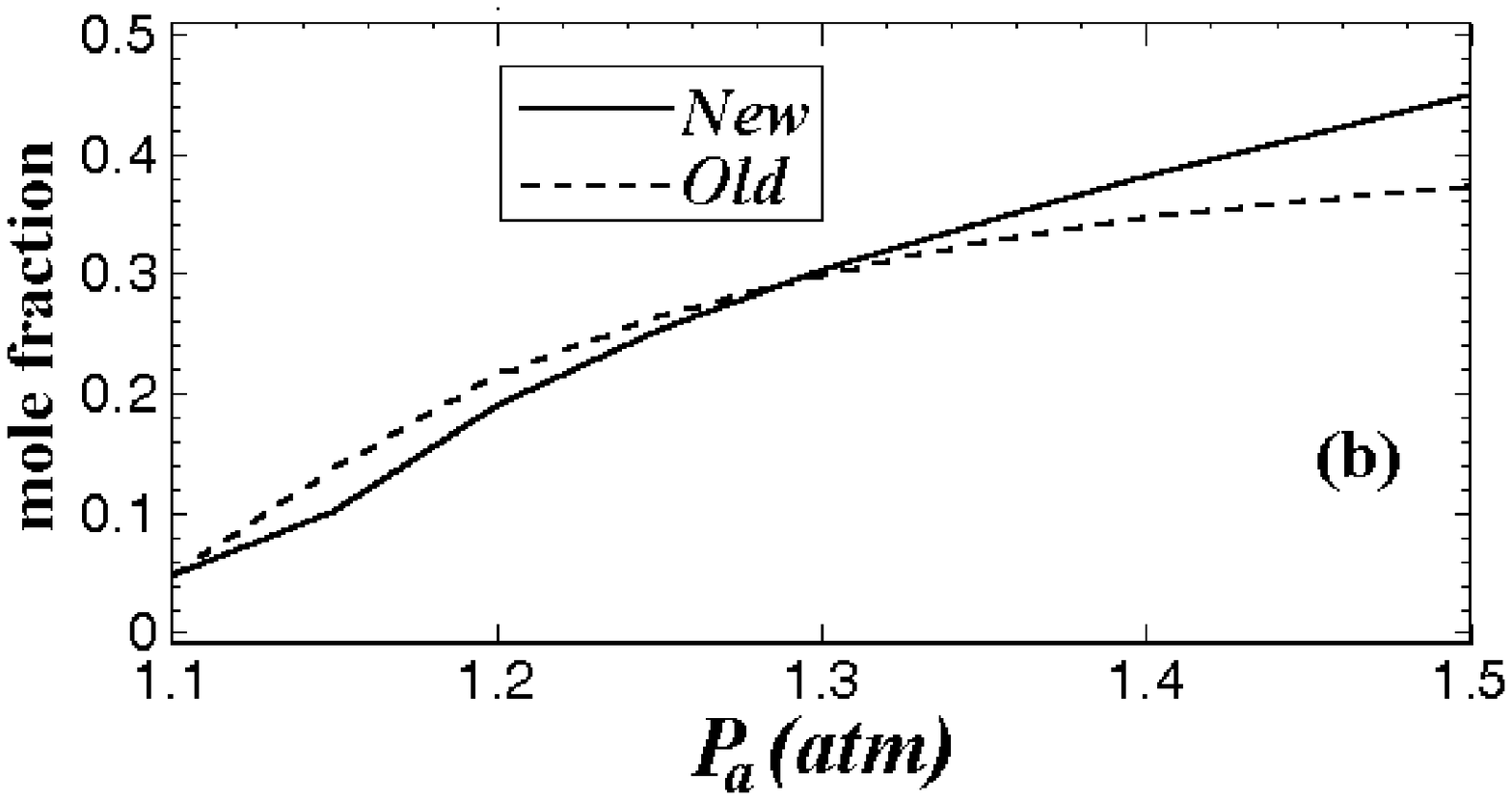}
\vskip 8mm
\includegraphics[width=7.6cm,height=3.7cm]{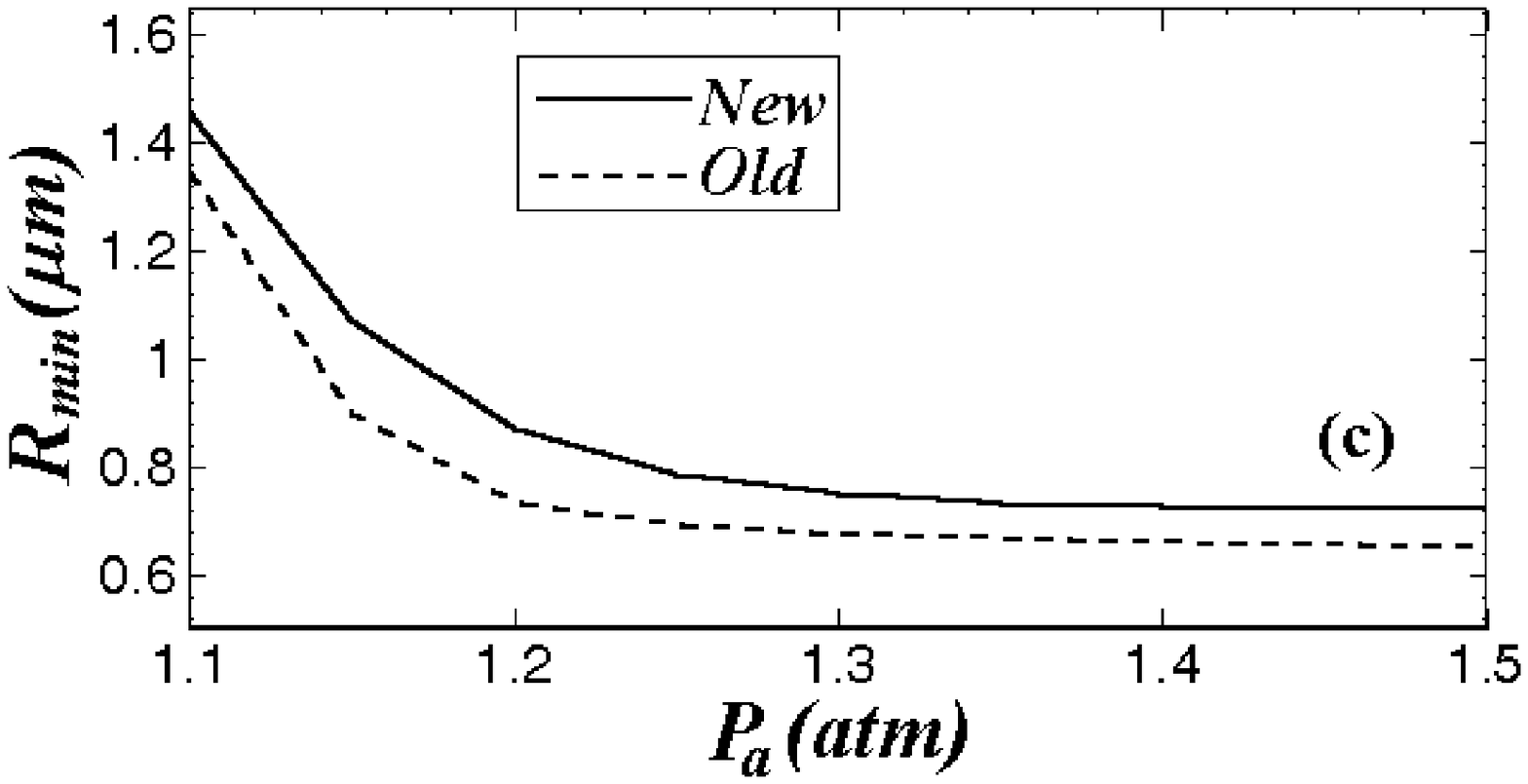}
\caption{The bubble characteristics at the time of collapse as a
function of driving pressure amplitude for the compressible
(solid) and incompressible (dashed) boundary conditions; peak
temperature (a), mole fraction of H$_2$O and reaction products
(b), and minimum radius (c). The equilibrium radius was fixed
($R_0=5.0 ~\mu m$). Other constants are the same as Figs. (1) and
(2).} \label{fig3:dls}
\end{figure}
Figure (3) represents the effects of variations of $P_a$ on the
bubble characteristics at the end of the collapse (peak
temperature, mole fraction of H$_2$O and reaction products, and
minimum radius), for the two boundary condition cases. The
ambient radius was fixed ($R_0=5.0 ~\mu m$). Figure 3(a) shows
that the peak temperature in the both cases grows as the driving
pressure is increased. However, the rate of increase of the peak
temperature for the new case is considerably smaller than that of
the old one. This causes that the difference between the two
cases becomes remarkable for the higher driving pressures (about
50\% for $P_a=1.5 ~atm$).

The bubble temperature at the end of the collapse is high enough
to destroy the chemical bonds of water vapor molecules inside the
bubble. The products of the dissociation of water vapor molecules
are mainly H$_2$, OH, H, O, and O$_2$. The chemical reactions
between the particle species existing inside the bubble affect
the bubble content at the collapse and its peak temperature
\cite{Kamath:1993,Yasui:1997,Storey:2000,Brenner:2002,Toegel:2002,
Toegel:2003, Lu:2003}. Here, we have considered the effects of the
reactions No. 1-8 of Refs. \cite{Yasui:1997, Toegel:2003}. The
dependence of the mole fraction of H$_2$O plus reactions
products, which is defined as ($N_{tot}-N_{Ar}$)/$N_{tot}$, to
the driving pressure for the two boundary condition cases has
been illustrated in Fig. 3(b). It shows that the mole fraction of
H$_2$O plus reactions products is similar for the two cases in low
amplitudes. The difference appears for the higher deriving
pressures. It is seen that, for the higher deriving pressures, the
effect of H$_2$O and reactions products is more important in the
new case relative to the old one.

Figure 3(c) shows the variations of the minimum radius as a
function of $P_a$, for the two cases. The trend of variations is
similar for the two cases. But, the minimum radius for the new
equation is more than that of the old equation because of the
reduction of the collapse intensity.

A major deficiency of the old bubble dynamics equations is that
for strongly driven bubbles, such as sonoluminescence bubbles,
large amplitude rebounds are produced after the collapse, so that
they often last until the next acoustic cycle of the periodic
driving pressure. This is in contrast with the experimental
results, which show rapidly damped rebounds \cite{Moss:2000}. By
introducing a damping term arisen from the gas compressibility,
Moss \textit{et. al} \cite{Moss:2000} provided a typical solution
for this problem. The effects of the suggested term by Moss
\textit{et. al} is very similar to the damping effects of the new
term in this paper, (compare Fig. 1(b) with Figs. (3) and (4) of
Ref. \cite{Moss:2000}). It seems that the damping feature of the
bulk viscosity is a better way for solving the mentioned problem.
The reason is that Eq'n. (\ref{eq9}) has been derived directly
from the basic equations of fluid mechanics, on the contrary to
Eq'n. (3.2) of Ref. \cite{Moss:2000}, which was derived by an
approximate method.

According to the results of this paper, it is expected that the
theoretical predictions of the bubble stability limits are
affected by the addition of the new term to the bubble dynamics
equations.

This work was supported by Sharif University of Technology and
Bonab Research Center. Partial support of this work by Institute
for Studies in Theoretical Physics and Mathematics is
appreciated. The authors thank Andrea Prosperetti for his helpful
comments.


\begin{thebibliography}{21}
\bibitem{Rayleigh:1917} L. Rayleigh, Philos. Mag., \textbf{34}, 94 (1917);
M. S. Plesset, J. Appl. Mech. \textbf{16}, 277 (1949).
\bibitem{Noltingk:1951} B. E. Noltingk and E. A. Neppiras, Proc. Phys. Soc. London B\textbf{ 63}, 674 (1950);
B. E. Noltingk and E. A. Neppiras, Proc. Phys. Soc. London B
\textbf{ 64}, 1032 (1951).
\bibitem{Herring:1941} C. Herring, OSRD Rep. No. \textbf{236} (NDRC C4-sr-10-010) (1941);
L. Trilling, J. Appl. Phys. \textbf{23}, 14 (1952); F. R. Gilmore,
Rep. No. \textbf{26-4}, Hydrodyn. Lab., Calif. Inst. Tech. (1952).
J. B. Keller and I. I. Kolodner, J. Appl. Phys. \textbf{27}, 1152
(1956). H. G. Flynn, J. Acoust. Soc. Am. \textbf{57}, 1379 (1975).
\bibitem{Keller:1980} J. B. Keller and M. Miksis, J. Acoust. Soc. Am. \textbf{68}, 628 (1980);
R. L\"{o}fstedt, B. P. Barber, and S. J. Putterman, Phys. Fluid A
\textbf{5}, 2911 (1993); R. I. Nigmatulin, I. SH. Akhatov, N. K.
Vakhitova, and R. T. Lahey, J. Fluid Mech. \textbf{414}, 47
(2000).
\bibitem{Prosperetti:1987} (a) A. Prosperetti and A. Lezzi, J. Fluid Mech. \textbf{168}, 457 (1986); (b) A Lezzi and A. Prosperetti, J. Fluid Mech. \textbf{185}, 289 (1987).
\bibitem{White:1991} F. M. White, \textit{Viscous Fluid Flow}, \textbf{2}nd edition (McGraw-Hill, New York, 1991), Chap. 2, p.67.
\bibitem{C.C.WU} C. C. Wu and P. H. Roberts, Phys. Rev. Lett. \textbf{70}, 3424 (1993).
\bibitem{Moss:1994} W. C. Moss, D. B. Clarke, J. W. White, and D. A. Young, Phys. Fluids \textbf{6}, 2979 (1994); W. C. Moss, D. B. Clark, and D. A. Young, Scince, \textbf{276}, 1398 (1997).
\bibitem{Kondict:1995} L. Kondic, J. I. Gersten, and C. Yuan, Phys. Rev. E \textbf{52}, 4976 (1995).
\bibitem{Voung:1996} V. Q. Voung and A. J. Szeri, Phys. Fluids \textbf{8}, 2354 (1996).
\bibitem{Yuan:1998} L. Yuan, et. al, Phys. Rev. E \textbf{57}, 4265 (1998).
\bibitem{Xu:2003} N. Xu, R. Apfel, A. Khong, X. Hu, and L. Wang, Phys. Rev. E \textbf{68}, 016309 (2003).
\bibitem{Kamath:1993} V. Kamath, A. Prosperetti, and F. Egolfopoulos, J. Acoust. Soc. Am. \textbf{94}, 248 (1993).
\bibitem{Yasui:1997} K. Yasui, Phys. Rev. E \textbf{56}, 6750 (1997).
\bibitem{Storey:2000} B. D. Storey and A. J. Szeri, Proc. Roy.
Soc. London, Ser. A  \textbf{456}, 1685 (2000);
\bibitem{Barber:1997} B. P. Barber, R. A. Hiller, R. L\"{o}fstedt, S. J. Putterman, and K. R. Weninger, Phys. Rep. \textbf{281}, 65 (1997).
\bibitem{Brenner:2002} M. P. Brenner, S. Hilgenfeldt, and D. Lohse, Rev. Mod. Phys.\textbf{74}, 425 (2002).
\bibitem{Toegel:2002} R. Toegel, S. Hilgenfeldt, and D. Lohse, Phys. Rev. Lett. \textbf{88}, 034301 (2002).
\bibitem{Toegel:2003} R. Toegel, D. Lohse, J. Chem. Phys. \textbf{118}, 1863 (2003).
\bibitem{Lu:2003} X. Lu, A. Prosperetti, R. Toegel, and D. Lohse, Phys. Rev. E. \textbf{67}, 056310 (2003).
\bibitem{CRC:1991} \textit{CRC Handbook of Chemistry and Physics},
edited by D. Lide, CRC Press, Boca Raton, (1995).
\bibitem{Karim:1952} S. M. Karim, L. Rosenhead, Rev. Mod.
Phys.\textbf{24}, 108 (1952).
\bibitem{Moss:2000} W. C. Moss, J. L. levatin, and A. J. Szeri, Proc. Roy. Soc. London, Ser. A
\textbf{456},2983 (2000).
\end{thebibliography}
\end{document}